# Informatics Carnot Machine

Oded Kafri

Varicom Communications, Tel Aviv 68165 Israel.

## Abstract

Based on Planck's blackbody equation it is argued that a single mode light pulse, with a large number of photons, carries one entropy unit. Similarly, an empty radiation mode carries no entropy. In this case, the calculated entropy that a coded sequence of light pulses is carrying is simply the Gibbs mixing entropy, which is identical to the logical Shannon information. This approach is supported by a demonstration that information transmission and amplification, by a sequence of light pulses in an optical fiber, is a classic Carnot machine comprising of two isothermals and two adiabatic. Therefore it is concluded that entropy under certain conditions is information.

A sequence of light pulses transmitted through an optical fiber is widely used in communication [1]. A random sequence of identical light pulses, representing "1" and vacancies, representing "0", is the physical entity of a transmitted binary file. The length $L$ of the pulse's sequence (the number of the pulses and vacancies) and the randomness of the distribution of the pulses determine the amount of the Shannon information being transmitted [2]. In this communication a thermodynamic analysis of the transmission of a random sequence of a light pulses (a file) is considered. It is shown that the Shannon information is entropy, and the amplification process done in an optical fiber is a Carnot cycle having the Carnot efficiency.

To calculate the entropy of the sequence of pulses, it is first necessary to calculate the entropy of a single, single-mode coherent pulse. It is assumed that the $i^{th}$ pulse in a sequence has $n_i$ photons, of energy $hv$, in a single mode. Since the photons are indistinguishable, the pulse is coherent. The temperature of the pulse will be assumed to be equal to that of a blackbody that emits $n_i$ photons into a single-mode of frequency $v$ [3]. Since a blackbody is in equilibrium with its radiation, a temperature can be calculated. (Appropriate spatial and spectral filters may filter the other radiation modes). In this case

$$n_i = \frac{1}{e^{hv/k_B T_i} - 1} \qquad (1)$$

The temperature of the coherent pulse obtained from eq. (1) to be

$$T_i = \frac{hv}{k_B \ln(1 + \frac{1}{n_i})}, \qquad (2)$$

The total energy of the pulse $q_i$ is $n_i h\nu$. Therefore the entropy that that a single pulse carries away from the blackbody is $S_i = q_i / T_i$, or:

$$S_i = n_i k_B \ln(1 + \frac{1}{n_i}) \qquad (3)$$

Since $\lim_{n_i \to \infty} S_i = k_B$ it means that the entropy of a coherent pulse is identical to that of a classic harmonic oscillator (namely $q_i = k_B T_i$) and is not a function of its energy. Similarly, a mode without energy carries no entropy as $\lim_{n_i \to 0} S_i = 0$.

The total entropy of a sequence of pulses, some with a large number of photons, (having entropy $k_B$), and some with no photons (empty pulses having no entropy), is the Gibbs mixing entropy of the sequence, namely, $S = -k_B \sum_{j=1}^{\Omega} p_j \ln p_j$. Where $\Omega$ is the number of configurations of the pulses and $p_j$ is the probability of the $j^{th}$ configuration. It is seen that the Shannon information and entropy are equivalent in the classical limit (a large $n_i$) to the Gibbs entropy of mixing.

For example, to calculate the entropy of a random sequence of light pulses, of length $L$, it is necessary to consider the fact that each pulse has a probability of ½ to be "one" and probability of ½ to be "zero". Therefore, the mixing entropy term is $S_i = -k_B(\frac{1}{2} \ln \frac{1}{2} + \frac{1}{2} \ln \frac{1}{2}) = k_B \ln 2$. To find the total entropy of the sequence we can sum all the entropies of the pulses (because the entropy is extensive), namely, $S = \sum_{i=1}^{L} S_i = k_B L \ln 2$. The Shannon information $I$ is defined as $I = -\sum_{j=1}^{\Omega} p_j \ln p_j$ [2]. If the probability of all the configurations is equal to $1/\Omega$ than $I = \log \Omega$. The number of the configurations of a binary file of a length $L$ is $2^L$, therefore the maximum amount of the information in the bits of a file having $L$ pulses is $L\ln2$ nats (1bit = ln2 nat). It is seen that thermodynamics and information theory yield the same result.

When the sequence of the bits is not random, the amount of information of the sequence is smaller. Therefore, in general, we obtain the Clausius inequality,

$$S \geq k_B I. \tag{4}$$

One can generalize this analysis and calculate the energy and the temperature of the whole sequence of pulses. This can be done easily for a random sequence. When $n_i$ is large, the temperature $T_i$ of a coherent pulse is $n_i h\nu/k_B = q_i/k_B$ where $q_i$ is the energy of the pulse. Namely, with contradistinction to the entropy, $T_i$ is a function of $q_i$. If we assume that all the energetic pulses have an equal energy $q$, the total energy of the sequence is $Q = \sum_{i=1}^{L} q_i = \frac{q}{2}L$. The entropy of the sequence is $S = k_B L \ln 2$ in nats or $k_B L$ in bits. The file temperature $T$ is $Q/S = q/2k_B$. This means that the average bit energy $q/2$ is equal to $k_B T$. This is the same relation as of a harmonic oscillator. It is worth noting that a random sequence of pulses is a non-coherent radiation, nevertheless it retains the thermodynamic properties of a harmonic oscillator.

Now it is shown that this formalism complies with the second law of thermodynamics [4]. Consider a long optical fiber in which a file comprising a sequence of light pulses, having a temperature $T_H$, with a pulse energy $q_H$, travels along the fiber. The pulse energy is attenuated due to the loss in the fiber. Therefore, the energy and the temperature of the pulse are reduced to $T_C$. Nevertheless, the amount of information (the entropy) remains intact. This process of cooling at constant entropy, thermodynamically speaking, is an adiabatic expansion. When the pulse energy reduces, the file requires amplification. To amplify the sequence of the pulses, the amplifier has to read the file first. The reading process is an energy transfer to the amplifier at constant bit energy. In this process the amplifier increases its energy at a constant temperature. Thermodynamically speaking, this process is an isothermal compression. In the next stage the file is amplified back to $T_H$. This stage is an adiabatic compression in which we invest work to increase the energy of the pulses without increasing their information (entropy). Finally, at the last stage the amplifier writes (emits) the light pulses into the fiber. In this stage the amplifier reduces its energy at a constant temperature $T_H$ and the cycle starts again.

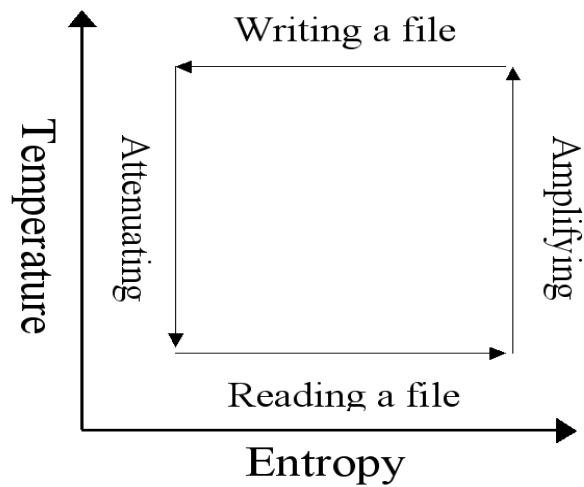

Fig. 1. A *Carnot cycle for a file amplification. Amplifiers are necessary to overcome the energy loss along a fiber. Each cycle of amplification is a Carnot cycle of two isothermals and two adiabatic.*

Hereafter it is shown that this cycle has an efficiency of the Carnot machine. Before entering the amplifier, the sequence of the light pulses has a relatively low temperature $T_C$, and its entropy is $Q/T_C$. After the amplification, it has a higher temperature $T_H$. If $Q$ is unchanged, the entropy is smaller. Therefore, the entropy balance $\Delta S = Q/T_H - Q/T_C < 0$ is negative, which is a violation of the 2$^{nd}$ law. The physical reason for the entropy reduction is that with a given amount of energy $Q$, one can write more low-energy bits than one can write high-energy bits. To conserve the entropy (a reversible operation), we have to keep $\Delta S = 0$. That means that we have to add more energy to $Q$. For a reversible operation $Q_H/T_H = Q_C/T_C$. Designating $Q_C = Q$ and $Q_H = Q+W$, we obtain, $W = Q(1 - T_C/T_H)$. In the irreversible case $Q_H/T_H > Q_C/T_C$, thus in general we obtain;

$$\eta \equiv \frac{W}{Q} \leq (1 - \frac{T_C}{T_H}) \qquad (5)$$

Eq. (5) is the Carnot efficiency.

Summary and discussion: The amount of entropy removed from a blackbody by a single radiation mode in the classical limit is $k_B$. If the radiation mode is empty it does not remove entropy. The entropy removed from the blackbody is assumed to be the entropy of the pulses. Therefore, it is argued that in the classical limit an energetic pulse carries $k_B$ entropy and a vacancy carries a zero entropy. When this assumption is used to calculate the entropy of a sequence of pulses, the obtained entropy of the sequence is the Gibbs mixing entropy, which is identical to the logical Shannon information. The plausibility of this formalism is demonstrated by presenting an informatics Carnot Cycle that yields the Carnot efficiency for an ideal amplifier cycle in an optical fiber.

Temperature and thermal equilibrium are concepts that are used to describe random systems in equilibrium. In random systems energy is exchanged between particles by collisions. There is no energy exchange between photons. Nevertheless, the quenched randomness of the energetic bits and the zero bits behaves according to the present formalism as in equilibrium, namely, a state where it is possible to calculate a unique temperature.

It was shown previously that laser operation [5,6] and laser-cooling processes [7] which involve a production or a usage of a coherent light, yield the Carnot efficiency, and therefore comply with the second law of thermodynamics. In these processes the light was considered as work, as light radiation was assumed to be coherent. A coherent light beam has a single radiation mode [8] and therefore it carries negligible amount of entropy. In the present study $L/2$ pluses, distributed randomly in $L$ modes, carry entropy that is shown to be the Shannon information. The pulse sequence is not coherent, as it is random. The lower the coherence, the higher is the amount of information that can be carried by the sequence. This communication suggests that the Shannon information can affect the efficiency of a Carnot machine.

Only when $n_i$ is large the entropy is not a function of the energy and the temperature. In this limit, the entropy becomes a pure measure of the quenched randomness, exactly as the logical Shannon information. This is a vital condition in informatics, as the entropy should remain intact with the energy attenuation. When $n_i$ is small, the entropy $S=S(Q)$ is smaller than that of the logical information. The entropy deficiency $k_B I - S(Q)$ is a loss of the logical information.

The Carnot efficiency of an amplifier can be tested experimentally. Calorimetric experiments of this kind require careful photon counting; nevertheless they are possible in the contemporary technology. This study suggests that the second law is applicable in the classical limit to informatics.

**Acknowledgements**: I thank Y.B. Band, R.D. Levine and Y. Kafri for many useful discussions.